\documentclass[runningheads,a4paper]{llncs}
\usepackage{amssymb} 
\usepackage{graphicx}
\usepackage{url}
\usepackage{float}
\usepackage{soul,color}  
\usepackage{amsmath}

\begin{document}

\title{The Internet of Hackable Things}

\author{
Nicola Dragoni\inst{1,2},
Alberto Giaretta\inst{2} and
Manuel Mazzara\inst{3}
\institute{DTU Compute, Technical University of Denmark, Denmark
\and Centre for Applied Autonomous Sensor Systems, \"{O}rebro University, Sweden
\and Innopolis University, Russian Federation}}

\toctitle{Lecture Notes in Computer Science}
\tocauthor{Authors' Instructions}
\maketitle

\begin{abstract}
The Internet of Things makes possible to connect each everyday object to the Internet, making computing pervasive like never before. From a security and privacy perspective, this tsunami of connectivity represents a disaster, which makes each object remotely hackable. We claim that, in order to tackle this issue, we need to address a new challenge in security: education.
\end{abstract}

\section{The IoT Tsunami}
\label{sec:iot_tsunami}

In the last decade, we all have witnessed a turmoil of interest around the Internet of Things (IoT) paradigm. It has been claimed that such a paradigm may revolution our daily lives and pervasive applications are behind the corner both in the civil and military complex. Such a strong hype on pervasive technologies requires a step back to consider the potential threat on security and privacy. First of all, What exactly is the IoT? Accordingly to the Online Oxford Dictionary it is the ``interconnection via the Internet of computing devices embedded in everyday objects, enabling them to send and receiving data''. To get a grasp of the dimension of this phenomenon, according to Evans Data Corporation the estimated population of IoT devices in June 2016 was 6.2 billion~\cite{E16}, number that according to several predictions will grow as up as 20 billion in 2020~\cite{GartIoT15}. Projections and data are not so straightforward to analyse since some firms take into account devices like smartphones, while others do not count them, therefore it is quite hard to make comparisons. Nonetheless, the growing trend is confirmed by every analyst, to the point that by 2025 the IoT market could be worth \$3.9 trillion to \$11 trillion per year~\cite{M15}. On the academic front, this ongoing excitement and interest in all the IoT world has given rise to an increasing number of related conferences, research projects and  research centres (like the recently formed IoT Center in Denmark, \url{http://iotcenter.dk}). 

\begin{figure}[ht]
\center
\includegraphics[trim=0mm 0mm 0mm 0mm,clip,width=0.75\textwidth]{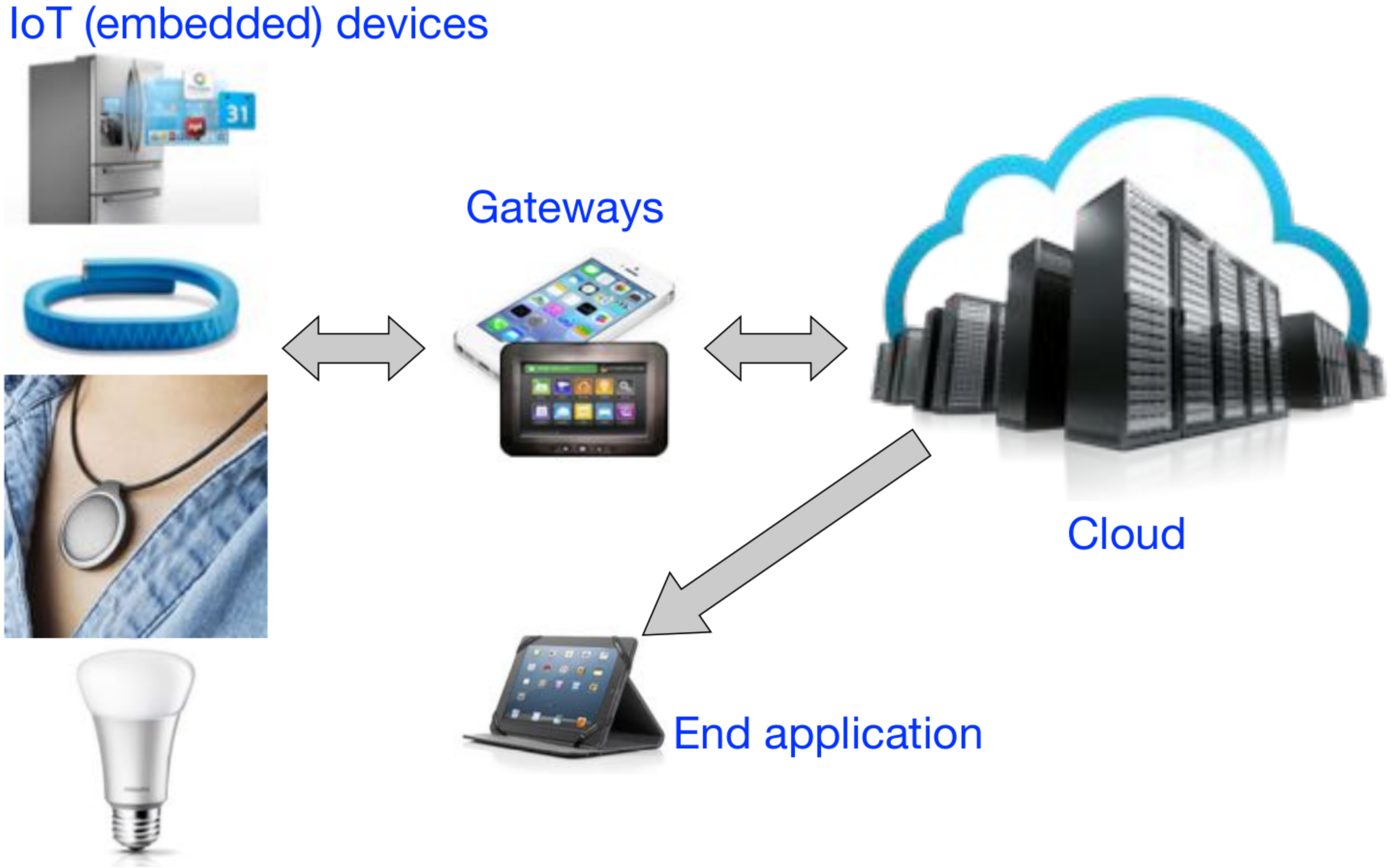}
\caption{Overview of a generic IoT architecture}
\label{fig:IoT_Arch}
\end{figure}

As a matter of fact, even though IoT refers to an ample variety of different devices, these devices all share a common architecture. First of all, any IoT device usually connects to the Internet through a more powerful gateway, which could be a smartphone or a tablet. Then data flow is elaborated by (and eventually hosted into) the cloud, enabling the end user to remotely connect to the device and control it. Figure~\ref{fig:IoT_Arch} shows how this IoT architecture looks like in a generic scenario.

IoT applications span from industrial automation to home area networks and personal (body) area networks. In particular, Smart homes will heavily rely upon IoT devices to monitor the house temperature, eventual gas leakages, malicious intrusions and several other parameters concerning the house and its inhabitants. Another growing area of interest is represented by pervasive healthcare applications, which use IoT devices to perform continuous biological monitoring, drug administration, elderly monitoring and so on. Last, but not least, in the recent years wearable devices gained a huge popularity (e.g., fitness trackers), to the point that in the span of just a year sales grew 18.4\% in 2016~\cite{GartWearable16}.


%
%
%
%

\subsection{A Security and Privacy Disaster}
\label{subsec:security_disaster}
From a security perspective, this ongoing excitement for IoT is having tremendous consequences, so that it's not an exaggeration to talk about a security and privacy disaster. Indeed, if the fundamental IoT axiom states that ``everything can be connected to the Internet (becoming, in this way, an IoT device)'', its security corollary is somehow catastrophic:  ``everything that can be connected to the Internet can be hacked"~\cite{P15}. This is particularly critical if we consider that, by means of the various kinds of devices connected to the Internet, people are sharing more and more information about themselves, often without being aware of that. This means that the amount of data available online is going to increase unrelentingly, literally given away to cybercriminal eager to take control of our devices, and thus of our life. In the early days of the ``IoT shift'', researchers highlighted how much critical security would be in a real IoT context~\cite{RNL11} and gave some hints about what should be done to defend our devices and our privacy. This message has clearly not been listened.

To put things in perspective, in July 2014 HP Security Research \cite{SAID15} analysed 10 of the most popular IoT devices on the market revealing a generally alarming situation:
\begin{itemize}
\item 90\% of devices collected at least some information via the device;
\item 80\% of devices, along with their cloud and mobile components, did non require a password complex enough;
\item 70\% of devices, along with their cloud and mobile components, enabled an attacker to identify valid user accounts through enumeration;
\item 70\% of devices used unencrypted network services;
\item 6 out of 10 devices that provided user interfaces were vulnerable to a range of weaknesses, such as persistent XSS\footnote{Cross site scripting (XSS) is an attack that injects malicious code into a Web application.} and weak credentials.
\end{itemize}
To make matters worse, security in a IoT scenario is even harder than expected for a number of reasons \cite{SITP14}, such as: \begin{itemize}
\item It implies complex and distributed systems, with a huge variety of different operating sistems, programming languages and hardware;
\item Even developing a simple application for a IoT device can be non-trivial;
\item Securing the applications is even less easy, because the attack surface is enormous (any device could be a possible entry point) and defining beforehand all the potential threats is extremely challenging;
\item The contained data are sensitive and highly valuable for the market, nowadays, which entails huge potential gains for any successful attacker and high attractiveness.
\end{itemize}

Given that providing security for the IoT is still a really hard thing to do, the atavistic problem with exciting new technologies is that companies are in a hurry and most of them ignore quite at all any kind of security issues, postponing the matter as much as possible. Just to give some numbers, Capgemini Consulting in 2015 highlighted some critical aspects~\cite{Cap15}, such as:
\begin{itemize}
\item Only 48\% of organizations focus on security of their devices from the beginning of the development phase;
\item Only 49\% of organizations provide remote updates for their devices;
\item Only 20\% hire IoT security experts;
\item Only 35\% invite third parties (like hackers) to identify vulnerabilities in their devices.
\end{itemize}

As a rule of thumb, we could depict the prevalent approach of manufacturers to IoT security with the following ``insecurity practice'' rule~\cite{SITP14}:
 \begin{equation}
 \begin{aligned}
             & Development~Rush~+~Hard~to~Develop\\
 \Rightarrow \quad & Skip~(or~Postpone)~Security
 \end{aligned}
 \end{equation}

At this point it should be quite easy to detect the reasons why hackers actually love the on-going IoT outburst. In the following Sections, we will show plenty of examples about this vast attention, with focus on two of the most promising IoT contexts: smart homes (Section~\ref{sec:home_horror}) and pervasive healthcare (Section~\ref{sec:perv_healthcare}).

\section{Smart Home... Of Horror!}
\label{sec:home_horror}
Smart homes and, in general, smart buildings are one of the current trends for IoT devices, and probably the most active one. 
Our team is also currently engaged in a project on microservice-based IoT for smart buildings~\cite{Salikhov2016a,Salikhov2016b}.
Everyday things are being transformed into much more powerful and smart objects, in order to meet customers' increasing needs. But availability of connected things could come with a high price in terms of privacy and security issues, in light of the fact that at the present moment too many things are too easily hackable.

Few years ago some irons imported from China included a wireless chip that was able to spread viruses by connecting to unprotected Wi-Fi networks, while some other hidden chips were able to use companies networks to spread spam on the Internet. Researchers achieved to hack the remote firmware update of a Canon Pixma printer, which makes possible to do funny things, like installing an old-school videogame such as Doom, and not so funny other ones, like installing a crippling malware that could even force the device to destroy itself.

Smart light bulbs, which enable the owners to remotely control and adjust their home light through an app or a web interface, are another fitting example of IoT devices. Some of these bulbs, such as the popular Philips Hues, have been compromised and researchers showed how easy is to set up a car, or even a drone, that drives in a residential area aiming to infect as much bulbs as possible with a crippling malware. This malware is able to shut them down or even force them to flicker on and off at desired speed~\cite{RFSW16}.

Smart TVs sales are constantly growing all over the world. Smart TVs provide a combination of a traditional TV and a Internet-connected personal computer, blending the two worlds into a single device. Usually these devices are equipped with various components, such as microphones and webcams, aiming to give the user the fullest experience possible. Clearly enough, if security is badly managed in these kind of devices, hackers could easily eavesdrop and peek at our lives without us even noticing that. An attack that could likely be struck is a HTML5 browser-based attack, therefore the devices resilience should always be assessed by using some penetration testing frameworks, such as BeEF~\cite{BeEF}.

Talking about spying, there are other devices that have been hacked with the specific intent to gather information about us. For instance, baby monitors are very unsafe devices, since that manufacturers generally equip them with default passwords easily guessable by attackers, passwords that usually are never changed by the customers. New York's Department of Consumer Affairs (DCA) issued a public statement~\cite{DCA} to inform people about the issue, even reporting that some parents walked in their child's room and heard some stranger speaking to them down the monitor.

Another perfect candidate to become a common IoT device in our smart home is the thermostat. Being able to remotely choose and monitor our house temperature can greatly benefit our wellness and comfort. Nonetheless, issues can arise too as shown by researchers at Black Hat USA, which demonstrated that a Nest thermostat (a popular device in the USA) could be hacked in less than 15 second if physically accessible by a hacker. The violated thermostat could be used to spy the residents, steal credentials and even infect other appliances. Recently, other researchers made a proof-of-concept ransomware that could remotely infect the aforementioned thermostat and shut down the heating, until the victim gives in to blackmail \cite{Thermostat}. Similar vulnerabilities have been found in many other smart home devices, where connectivity has been ``embedded'' in the device without considering any  security protection. 

Even more serious is the threat posed by the lack of security in top-selling home alarm systems, which unveiled weaknesses are critical to such an extent that a malicious attacker could easily control the whole system, suppressing the alarms or creating multiple false alarms. In fact, some of these systems do not encrypt nor authenticate the signals sent from the sensors to the control panel, easily enabling a third party to manipulate the data flow. 

Life-threatening vulnerabilities have been found even in smart cars. Security researchers at Keen Security Lab were able to hack a Tesla Model S, achieving to disrupt from a distance of 12 miles various electronically controlled features of the car, such as the brakes, the door locks and the dashboard computer screen \cite{Tesla}.

Last but not least, we have seen a proliferation of wearable health trackers in the last couple of years. In order to provide the user its monitoring features, a fitness tracker is an embedded system which collects sensitive data about the wearer and communicates it to a mobile application by means of a Bluetooth Low Energy (BLE) protocol, hence enabling the user to access the gathered information. Moreover, nowadays most of the mobile applications sync the collected data to a cloud service, whenever an Internet connection is available (see Figure \ref{fig:IoT_Arch}). Researchers conducted some deeper investigations about this whole system~\cite{GDS16}, evaluating the security of the implemented protocols in two of the most popular fitness trackers on the market. The research highlighted how vulnerable these devices are to several kinds of attacks, from Denial of Service (DoS) attacks that can prevent the devices from correctly working, to Man-In-The-Middle (MITM) attacks based on two fake certificates~\cite{CDG13} resulting in a disclosure of sensitive data. Worryingly, the implemented attacks can be struck by any consumer-level device equipped with just bluetooth and Wi-Fi capabilities (no advanced hacking tools have been required).

If you think that escaping from a hacked smart home to find some peace in a hotel room is a temporary solution, well you might be wrong. Recently, guests of a top-level hotel in Austria were locked in or out of their rooms because of a ransomware that hit the hotel's IT system. The hotel had no choice left except paying the attackers.

\section{Pervasive Healthcare}
\label{sec:perv_healthcare}

If the so-far depicted Smart Home scenario is already scary, things can even get worse when we look at the pervasive healthcare context, for example the the infrastructure to support elderlies developed by our team~\cite{Nalin2016}. Indeed, when we talk about security in healthcare we inherently talk about safety, since malfunctioning, attacks and lack of service could endanger many lives, as we will show in the following. 

\subsection{eHealth: How to Remotely Get Big Data}
\label{subsec:ehealth}
Duo Security highlighted how security is badly managed in healthcare corporations, showing that the density of Windows XP computers is 4 times greater than the density of machines running the same OS found, for instance, in finance. Given that Microsoft ended the support to Windows XP since 2014, this means that an enormous quantity of devices has not been updated for 2 years, at least. Not only obsolete operating systems, even additional (and most of the times, useless) software can become a problem: many healthcare endpoints and healthcare customers' terminals have Flash and Java installed, entailing a huge risk of vulnerabilities exploitation.

To get an idea of how much valuable eHealth data is, and consequently how critical the related security is, the global information service Experian estimated that on the black market health records are worth up to 10 times more than credit card numbers. Particularly, a single eHealth record (which comprises social security number, address, kids, jobs and so on) can be priced as high as \$500.

For the sake of clarity, we are definitely talking about risks which are far from theoretical: healthcare industry suffers estimated costs of \$5.6 billion per single year because of data thefts and systems malfunctioning. According to \cite{Anthem}, in February 2015 78.8 million of Anthem customers were hacked. In the same year, according to the Office of Civil Rights (OCR), more than 113 million medical records were compromised. Earlier last year Melbourne Health's networks got infected with a malware capable of keylogging and stealing passwords. In February 2016 Hollywood Presbyterian Medical Centre was struck with a devastating ransomware, conveyed by simple Word document in an email attachment. The most recent demonstration of hackers interest about eHealth data is a massive sale of patients records on the dark web, where more than 650.000 tags were auctioned off to the highest bidder.

What strikes the most is that we are dealing with a huge amount of data weakly defended, easily accessible and highly valuable to malicious third parties. People tend to link security to tangible money stored in bank accounts, but we've witnessed a radical shift about what's valuable in the black market, in the last decade. Hackers do not just want our credit cards, they want the patterns of our life.

\subsection{IoT Medical Devices: How to Remotely Kill You}
\label{subsec:med_devices}
The IoT revolution is particularly relevant for a number of healthcare fields of application, since networked devices make possible to monitor and deliver necessary treatments to any remote patient, meaning that day-to-day and even life-saving procedures can be promptly performed. Nowadays, devices like insulin pumps, cochlear implants and cardiac defibrillators are used on a daily basis to deliver remote assistance to a lot of patients. Furthermore, in the last years bigger devices like blood refrigeration units, CT scan systems and X-ray systems are connected to the Internet, in order to check remotely their operational state and make whatever adjustment is needed (e.g., lower the blood unit inside temperature).

\begin{figure}[ht]
\center\includegraphics[trim=0mm 0mm 0mm 0mm,clip,width=0.75\textwidth]{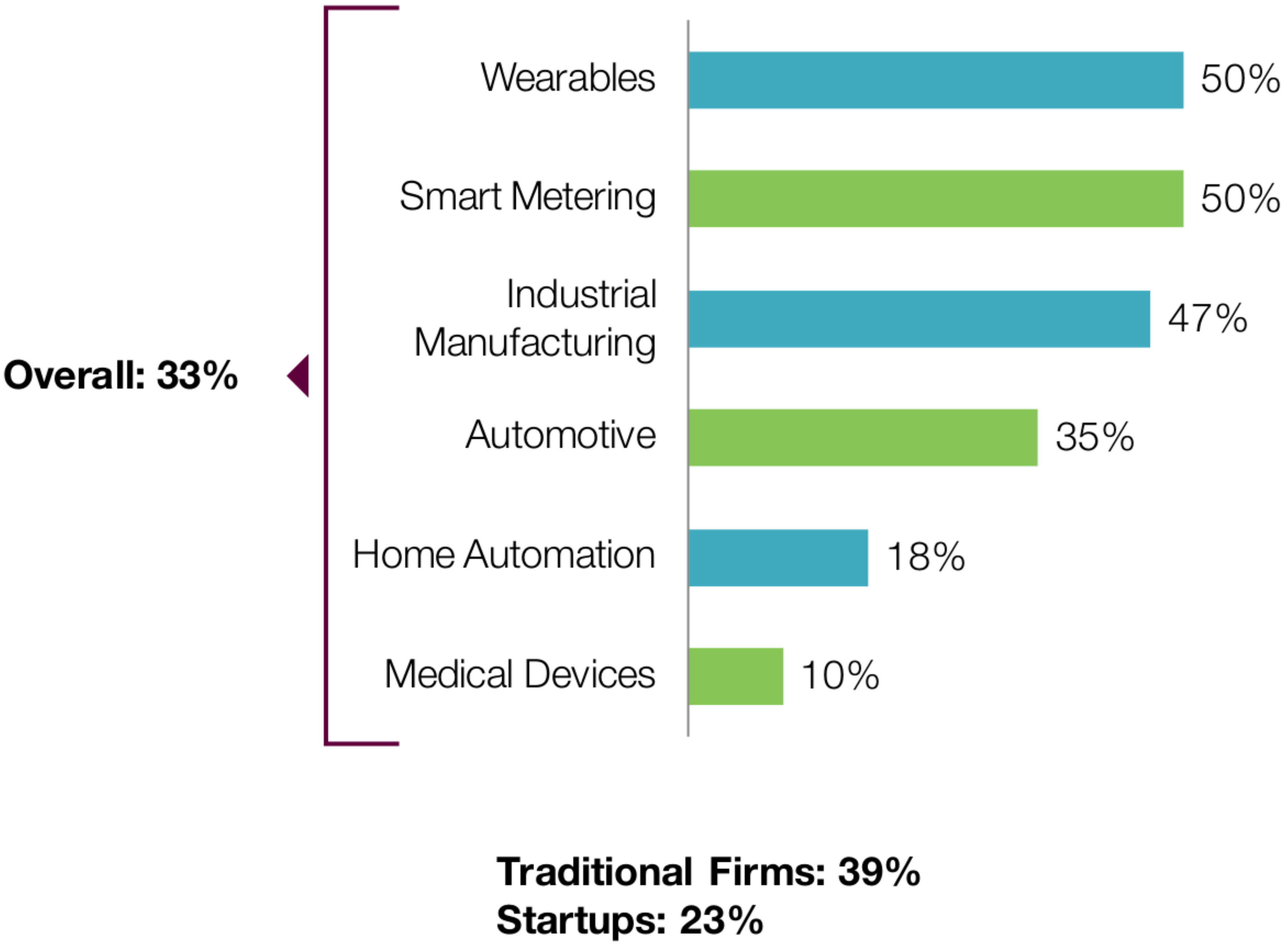}
\caption{Percentage of firms executives that rate the IoT products, in their own industry, highly resilient to cyber attacks~\cite{Cap15}.}
\label{fig:Segments_Resilience}
\end{figure}

Keeping in mind that, as we stated in Section~\ref{sec:iot_tsunami}, when something is connected to the Internet it is inherently not secure, the other side of the coin is that the IoT-based healthcare exposes the aforementioned life-saving procedures to the public domain. Therefore, this exposure entails that ``if it isn't secure, it isn't safe"~\cite{NB16}. For the sake of clarity, Capgemini Consulting conducted an investigation in February 2015~\cite{Cap15} where firms executives were asked about the resilience of IoT products in general, in their own opinion. Results shown in Figure~\ref{fig:Segments_Resilience} show that medical devices are critically at the bottom of the survey, with only a 10\% of executives that believe that IoT devices are highly resilient to cybercriminals. Indeed, various life-threatening vulnerabilities have been found in a number of IoT devices. At least 5 models of intravenous drug pumps manufactured by Hospira, an Illinois firm that administers more than 400.000 devices all over the world, recently showed critical vulnerabilities that could allow a malicious attacker to alter the amount of drugs delivery to patients. Medtronic, one of the world's largest standalone medical technology development company, manufactures an insulin pump that enables patients to autonomously manage their blood glucose levels; sadly, the system does not encrypt the commands sent to the pumps by patients, nor do authenticate the legitimacy of the user. Such an uncontrolled system means that unauthorized third parties could intercept a legitimate command and replace it, delivering a deadly insulin dose to the patient. Some companies that produce Implantable Cardioverter Defibrillators (ICDs), used to deliver shocks to patients going into cardiac arrest, use a Bluetooth stack to test their devices after the first implantation, but they use default and weak passwords which makes their product easily hackable. 

Similar problems have been found in blood refrigeration units, protected only by a hardcoded password that could be deciphered by malicious attackers and used to alter the refrigeration unit temperature, consequently wrecking the blood provision. 

Another attack could be struck by targeting CT scanning equipments and altering the radiation exposure limits, killing a patient by administering a huge amount of radiation. Even some X-ray systems have been proved to be vulnerable, as they do not provide any kind of authentication when patients' X-rays are backed up in centralized storage units, nor log who views the images.

Bad security can be as dangerous as lack of security, as seen in May 2016 when Merge Hemo, a medical equipment used to supervise hearth catheterization procedures, crashed due to a scan triggered by the antivirus software installed: installing antivirus e antimalware software is not only insufficient, sometimes it can even be hazardous if superficially done.

\section{On the Need of Developing a Security Culture}
\label{sec:need_culture}

Today technology is so sophisticated that counteracting outside threats requires a high level of knowledge and a vast set of skills. This becomes even more challenging if security is mostly unheeded as it happens today, treated as a postponable aspect of a product instead than a inherent and essential trait. And while firms struggle to keep on track, hackers keep on gaining competence and resources: as an example, ransomware victims receive easy and detailed instructions on how to unlock their devices, and sometimes hackers themselves provide 24/7 call centres, in case their targets should run into any kind of technical difficulty. Shockingly, the support victims get from hackers is better than the support they get from their own Internet Service Provider.

So, what are the recommendations that should be followed in designing more secure IoT devices? How can we mitigate, if not solving, this security and privacy disaster? 

We believe that, to provide an answer, we first need to step back to the basic question: what is the nature of the problem? Is it technological? Rephrasing, do we have a lack of proper technology to protect IoT systems? Do we need new security solutions? 

Our (probably provocative) answer is no, we do not need technological innovation. Or better, of course we do need that, as we do need government regulation, but these are not the priority. The priority is instead education. Indeed, what we actually miss is to develop an effective security culture, raising the levels of awareness and understanding of the cyber risk and embedding ``security-aware'' values and behaviours in our everyday life. 

Security and trust are indeed also matter of education and method. For example, in social networks algorithm to compute users trust exist~\cite{MBGDMQN2013}, still people need to rely on their own experience and understanding and should not blindly follow computer suggestions. It is the integration of human understanding and algorithms that always offer the best solutions.
 
To support the above argument, consider all the examples of IoT devices mentioned in this paper (a summary is given in Table \ref{tab:summary}). It is noteworthy to highlight that all the described vulnerabilities have the common characteristic of being possible thanks to the naive approach that manufactures adopted in the design phase of their products, approach that clearly shows how security is merely sketched out or even neglected at all. Following basic and well known security practices, it would have been possible to protect these devices against all those cyber-attacks. This is something extremely important to understand. For instance, just to provide another example supporting our argument, let us consider the Mirai malware that operated in October 2016, achieving the largest Distributed Denial of Service (DDoS) attack ever, approximately hitting the targets with 1.2~Tbps of requests \cite{INSERT17}. Mirai simply scans the Internet, looking for vulnerable IoT devices to attack with a simple dictionary approach and, once that access is gained, the device becomes a bot of a huge network ready to strike a massive DDoS attack. Noticeably, the dictionary used by Mirai is filled with a tiny number of entries, around 50 combinations of username/password, which gives an idea of how little effort is put by firms into designing security for their IoT devices, at the present moment. Again, what was the key issue making this huge cyber attack possible? Was it a lack of technological innovation, for instance a stronger authentication mechanism? Or a lack of a basic security culture, so that we do not apply the technology we already have and that could actually solve most of nowadays security vulnerabilities? 

\begin{table}[t]
\caption{Examples of hacked IoT devices. ``Weak security'' means that the device was easily breakable because of a lack of basic security protection mechanisms (details in the paper).}
\begin{center}
\begin{tabular}{|l|c|}
\hline
\textbf{IoT Device} & \textbf{Why Hacked} \\
\hline
Tea kettles & No security \\
Irons & No security \\
Kitchen appliances & No security \\
Printers & Weak security \\
Networked light bulbs & Weak security \\
Smart TVs & Weak security \\
Baby monitors & Weak security \\
Webcams & Weak security \\
Thermostats & No security \\
VoIP phones & Weak security \\
Home alarm systems & No security \\
Smart toilets & No security \\
Smart cars & Weak security \\
Drug infusion pumps & Weak security \\
Insulin pumps & No security \\
Implantable cardioverter defibrillators & Weak security \\
X-Ray systems & No security \\
Blood refrigeration units & Weak security \\
CT scanning equipment & No security \\
Heart surgery monitoring device & Weak security \\
Fitness trackers & Weak security \\
Hotel room doors & Weak security\\
\hline 
\end{tabular}
\end{center}
\label{tab:summary}
\end{table}%

Security best practices recommend that a detailed risk analysis should be done, in order to have a clear view of what are the actual cyber threats and consequently choose the right approach to secure the devices. Moreover, device security should be designed as an essential part of the product lifecycle and not as a one-time issue. Once that the right path has been chosen for the new products, already existing devices should be thoroughly tested, following a fairly simple schedule like: automated scanning of web interfaces, reviewing of network traffic, reviewing the need of physical ports (e.g., USB ports), reviewing authentication and authorization processes, reviewing the interaction of devices with cloud and mobile application counterparts (an example for health trackers is given in \cite{GDS16}).

In the end, what we have learned by this excursus is that the main problem and concern with IoT security is that a security culture is nearly non-existent in our society. It should sound obvious that the more the technology develops and becomes pervasive in our lives, the more the security awareness should be growing. But this is not happening, or it is happening at a too slow pace. Indeed, while the concept of ``computing'' has rapidly and significantly evolved in the last decades (from mainframes to personal computing to mobile and then pervasive computing), the development of security has not followed the same evolution. Nowadays, kids are able to use almost any mobile device like smart phones, laptops, tablets, wearable devices and so on. On the other hand, they have no concept of ``security'' or ``privacy''. With the explosion of IoT, computing has become pervasive like never before. It's time that also security becomes so pervasive, starting from the development of a new security culture. This is surely a long term goal that has several dimensions: developers must be educated to adopt the best practices for securing their IoT devices within the particular application domain; the general public must be educated to take security seriously, too, which among other things will fix the problem of not changing default password. This education effort, however, will surely need the support of both innovation and government regulations, in order to enforce security when education is not enough. 

We are strongly convinced that education is the key to tackle a significant number of today IoT security flaws. Therefore, if we raise the levels of cyber risks understanding, both in the corporations and in the general end-users, maybe what future holds would not look as daunting as it looks today. We call the research community to this new exciting challenge.

\end{document}